\documentclass[info]{epl}
\title{Magnetic resonance at 41 meV and charge dynamics in 
\chem{YBa_2Cu_3O_{6.95}}}
\shorttitle{Magnetic resonance at 41 meV \ldots}
\author{E. Schachinger\inst{1} \and J.P. Carbotte\inst{2} \and
D.N. Basov\inst{3}}
\institute{
\inst{1} Institut f\"ur Theoretische Physik, Technische Universit\"at
Graz, A-8010 Graz, Austria\\
\inst{2} Department of Physics and Astronomy, McMaster University,
Hamilton, Ont. L8S 4M1, Canada\\
\inst{3} Department of Physics, University of California, San Diego;
La Jolla, Cal 92093-0319
}
\begin{document}
\maketitle
\begin{abstract}
We report an Eliashberg analysis of the electron dynamics in
\chem{YBa_2Cu_3O_{6.95}}. The magnetic resonance at 41 meV
couples to charge carriers and defines the characteristic
shape in energy of the scattering rate $\tau^{-1}(T,\omega)$,
which allows us to construct the charge-spin spectral density
$I^2\chi(\omega,T)$ at temperature $T$. The $T$ dependence
of the weight under the resonance peak in $I^2\chi(\omega)$ agrees
with experiment as does that
of the London penetration depth and of the microwave conductivity.
Also, the $T=0$ condensation energy,
the fractional oscillator strength in the condensate, and the ratio of
gap to critical temperature agree well with the data.
\end{abstract}
A hallmark of the spin dynamics of  several classes of high-T$_c$
superconductors is a magnetic resonance observed around
$40\,$meV by means of spin polarized inelastic neutron scattering.
The position in energy of the peak scales with
the critical temperature both in YBa$_2$Cu$_3$O$_{6.95}$ (Y123) and
Bi$_2$Sr$_2$CaCu$_2$O$_8$ (Bi2212) superconductors \cite{He}. 
The analysis of the optical  conductivity shows that the charge carriers
are strongly coupled to the magnetic excitations with a coupling strength
sufficient to account for superconducting transition temperatures
$\simeq 90\,$K \cite{Carb1}. Thus, neutron scattering data combined
with optical conductivity results, signal the prominence of the resonance
mode for superconductivity in the cuprates. The analysis of additional
optical data by Schachinger and Carbotte \cite{Schach2} 
showed that similar magnetic resonances are expected to be found in many 
other high-$T_c$ materials with $T_c\simeq 90\,$K and that this phenomenon
is not restricted to bilayer materials. At the moment some of these 
cannot be investigated with neutrons due to minuscule crystal size.

Because the spectral weight under the resonance is maximized at $T\ll T_c$ 
and vanishes at $T\simeq T_c$ \cite{Dai} (at least in optimally doped Bi2212
and Y123), coupling of the conducting carriers to the magnetic mode ought
to influence the temperature dependence of a variety of properties including
the London penetration depth and microwave absorption. In this work we
determine the charge-spin spectral density $I^2\chi(\omega)$ at various
$T$ from the inversion of the optical constants. The $T$
dependence of the spectral weight under the resonance we obtain
matches well the neutron data \cite{Dai} on the temperature evolution of the 
magnetic resonance. We employ Eliashberg formalism to 
calculate the $T$  dependence of the  penetration depth, of the 
microwave conductivity and the fractional optical oscillator
strength that condenses in the superfluid density at $T=0$. These results
along with calculations of the zero temperature condensation energy and the
ratio of gap to critical temperature agree with
experiment. This establishes the pivotal role played by the
magnetic resonance mode in the charge dynamics as well as in the thermodynamic
properties of the cuprates. 

In a previous publication \cite{Carb1} we showed that an appropriately 
defined second derivative \cite{Mars1} of the optical scattering rate 
$\tau^{-1}(\omega)$ as a function of frequency gives an absolute measure
of the spin excitation spectrum weighted by their coupling to the charge
carriers. While the second derivative technique is not exact and, for
instance, does not account for the anisotropy of the relaxation rate on
different segments of the Fermi surface \cite{Rice}, it,
nevertheless, allows one to construct a good first estimate of the frequency
dependence of the underlying charge carrier-spin excitation spectral density.
Here we apply this analysis to study the temperature dependence
of the spectral weight under the spin resonance in optimally doped Y123.

Eliashberg theory, on the other hand, is a highly successful extension
of the BCS model
for the case of an electron-phonon mechanism \cite{Carb2}. While
the theory was formulated for the electron-phonon case, it can
be used as a first approximation for other boson exchange
mechanisms, the main limitation being that vertex corrections
may become more important and therefore the theory less
accurate. With the symmetry of the gap ($d$-wave) explicitely
introduced into the formalism, the only material parameter
that enters is the electron-spin excitation spectral density
$I^2\chi(\omega)$.

\begin{figure}
\onefigure[width=9cm]{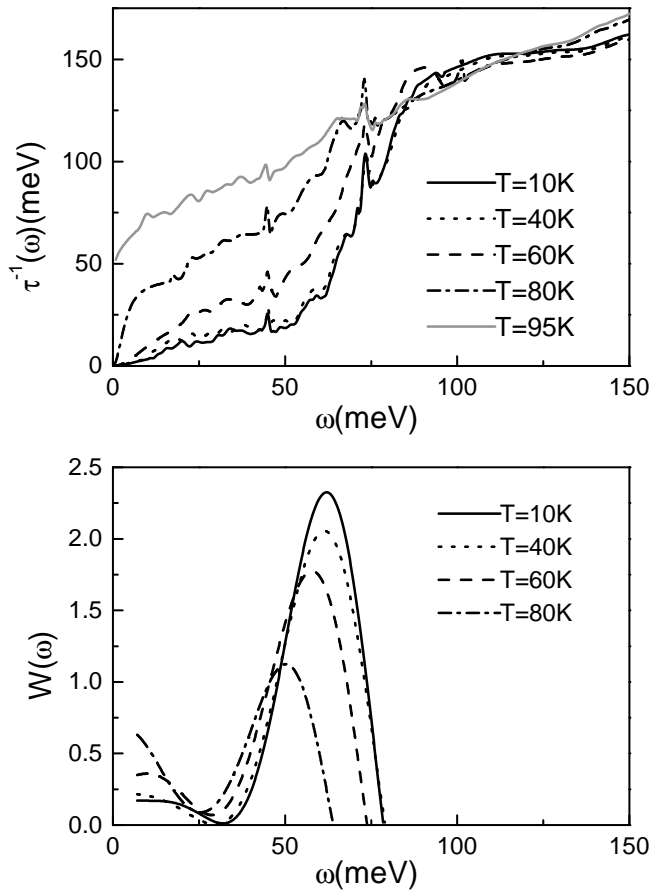}
\caption{Top frame: optical scattering rate $\tau^{-1}(T,\omega)$
in meV for optimally doped twinned Y123 single crystals.
Bottom frame: function $W(\omega)$ vs. $\omega$ in the region of
the spin resonance.}
\label{f1}
\end{figure}

\begin{figure}
\onefigure[width=9cm]{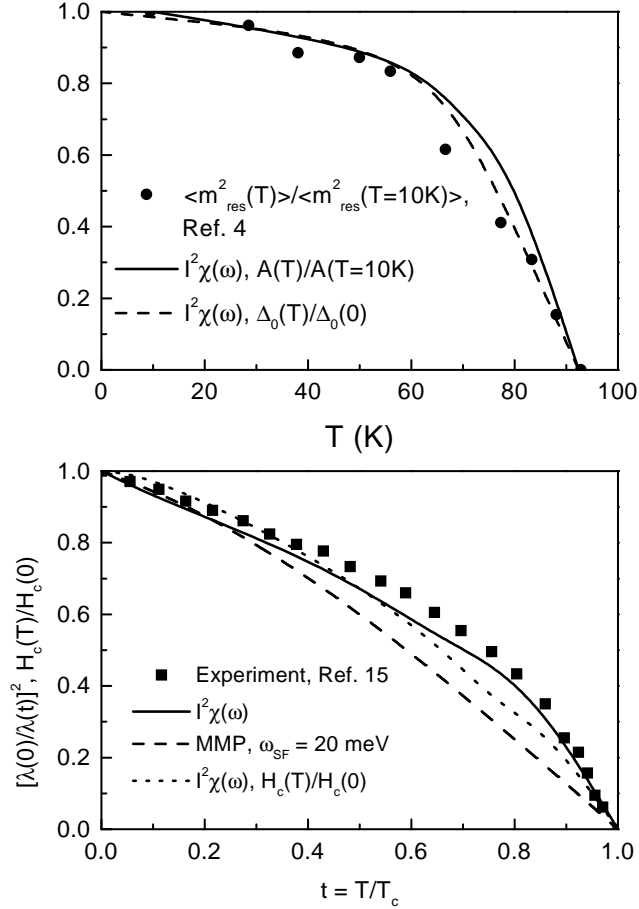}
\caption{Top frame: spectral weight under the spin resonance as
a function of temperature (solid line) obtained from the optical data
of Fig.~\ref{f1} (top frame) (5 temperatures only). The solid circles
are the data of
Dai {\it et al.} \cite{Dai} obtained by neutron scattering. The
dashed curve gives our calculated $\Delta_0(T)/\Delta_0(0)$.
Bottom frame: the normalized London penetration depth squared
$(\lambda(0)/\lambda(T))^2$ vs. reduced temperature $t = T/T_c$
(solid line) compared with
the experimental results of Bonn {\it et al.} \cite{Bonn1}. The
dashed line does not include the spin resonance and is for
comparison. The dotted curve gives
thermodynamic critical field $H_c(T)/H_c(0)$ vs.\ $t$.}
\label{f3}
\end{figure}

In Fig.~\ref{f1} (top frame) we present the raw experimental results for the
optical scattering rate spectra $\tau^{-1}(\omega)$ obtained for twinned
samples of optimally doped YBCO single crystals \cite{Basovn}. Here
$\tau^{-1}(\omega) = \Omega^2_p\Re{\rm e}\{\sigma^{-1}(\omega)\}/4\pi$
where $\sigma(\omega)$ is the infrared conductivity and $\Omega_p$ is
the plasma frequency. The low-$T$ spectra reveal a threshold structure
starting around $\omega \simeq 60\,$meV which is a common 
signature of the electromagnetic response of the CuO$_2$ planes in a 
variety of high-$T_c$ superconductors \cite{Puch}. With increasing $T$ this 
feature weakens and at $T>T_c$ the scattering rate assumes a nearly linear 
$\omega$-dependence which is the counterpart of the linear 
resistivity, and is characteristic of the marginal
Fermi liquid \cite{Varma}. The second derivative 
technique \cite{Carb1,Schach2} shows more clearly the threshold
structure and helps to unravel its microscopic origin.
This technique applies to a $d$-wave superconductor which is
described by a charge carrier-spin excitation spectral density 
$I^2\chi(\omega)$ within a generalized Eliashberg formalism 
\cite{Schach3,Schach4}. A peak in the spectral density
at energy $\omega_{sr}$ of the spin resonance, gives a peak at
$\omega\simeq \Delta + \omega_{sr}$ (where $\Delta$ is the energy gap) in the 
second derivative of $\omega\tau^{-1}(\omega)$. A first approximation to
the spectral density which applies however only to the region of the
resonance peak (beyond it $W(\omega)$ has a negative region, not part
of the spectral density which is positive definite) is given by the
function $W(\omega)/2$ with  
$\omega$ appropriately shifted by $\Delta$, and \cite{Mars1}
 \begin{equation}
  W(\omega) = {1\over 2\pi}{d^2\over d\omega^2}\left[
   {\omega\over\tau(\omega)}\right].
  \label{eq:1}
\end{equation}
The spectra for $W(\omega)$ (bottom frame of Fig.~\ref{f1}) indeed reveal a
peak at $\simeq 70$ meV at 10 K. Also, with increasing $T$, this peak softens
by about 20 meV but this shift coincides with the temperature
dependence of the gap, so that the energy of
the spin resonance is temperature independent and the spectral weight 
confined under the peak is gradually reduced to zero as $T$ approaches
$T_c$, and shows a
temperature dependence similar to the strength of the spin
resonance (top frame of Fig.~\ref{f3}). These observations suggest that the
spin resonance is responsible for the rise in the optical scattering rate.
Recent ARPES data \cite{Kaminski} show quasiparticle peaks along the
BZ-diagonal (most relevant to in-plane transport) with width that
display similar behaviour to the optical rates. 

The data shown in the bottom frame of Fig.~\ref{f1} can be used to construct
a charge carrier-spin excitation spectral density according to the following
prescription. It is found that the scattering rate shown at $T=95\,$K
can be fit with the form \cite{Millis1}
\begin{equation}
 I^2\chi(\omega) = G{\omega/\omega_{SF}\over 1+
  \left(\omega/\omega_{SF}\right)^2},
 \label{eq:3}
\end{equation}
where the single spin fluctuation energy $\omega_{SF} = 20\,$meV
produces a good fit to the $\omega$ dependence of the $T=95\,$K
data for $\tau^{-1}(\omega)$ with $G$ adjusted to get the correct
magnitude when a cutoff of $400\,$meV is applied. In the superconducting state 
the spectrum of $I^2\chi(\omega)$ given by Eq.~(\ref{eq:3}) is modified
only at small $\omega$ \cite{Carb1,Schach2} with
the spin resonance at $41\,$meV added according to the data given
in the lower frame of Fig.~\ref{f1}.
The area under the main resonance peak in
$I^2\chi(\omega)$ is defined as $A(T)$ and the normalized area
$A(T)/A(10\,{\rm K})$ is plotted in the top frame of Fig.~\ref{f3} as the
solid line. This compares well with the solid circles for the
normalized area under the neutron resonant peak obtained by Dai
{\it et al.} \cite{Dai} denoted by $\langle m^2_{\rm res}(T)%
\rangle/\langle m^2_{\rm res}T=10\,{\rm K}\rangle$. The agreement is
good and shows that the variation in neutron peak intensity with $T$
is reflected accurately in the transport data.

Once a model spectral density $I^2\chi(\omega)$ is specified 
superconducting properties follow from the solution of the $d$-wave 
Eliashberg equations \cite{Schach3,Schach4}. Here we report on two
such properties. In the bottom frame of Fig.~\ref{f3} we show
our results for the normalized 
London penetration depth $ [\lambda(0)/\lambda(t)]^2$ as a function of 
reduced temperature $t$ and compare with experimental results
obtained by Bonn {\it et al.} \cite{Bonn1} (solid squares).
The dashed line was obtained with a
spectral density taken as temperature
independent and fixed to its $T = T_c$ value; calculation details are given in
Ref.~\cite{Schach3}. The agreement with the data is poor
but can be significantly improved if we use the model for
$I^2\chi(\omega)$ which includes the spin resonance. The solid line
reproduces the essential
features of the experimental data. It is the growth in strength with
decreasing $T$ of the $41\,$meV peak and the loss of spectral weight
at small $\omega$ in the superconducting state that accounts for the bulging
upward of the solid curve in the region above $t \simeq 0.3$.

Another quantity of interest is the temperature dependence of the
microwave conductivity below $T_c$. A large peak \cite{Bonn1}
\begin{figure}
\onefigure[width=10cm]{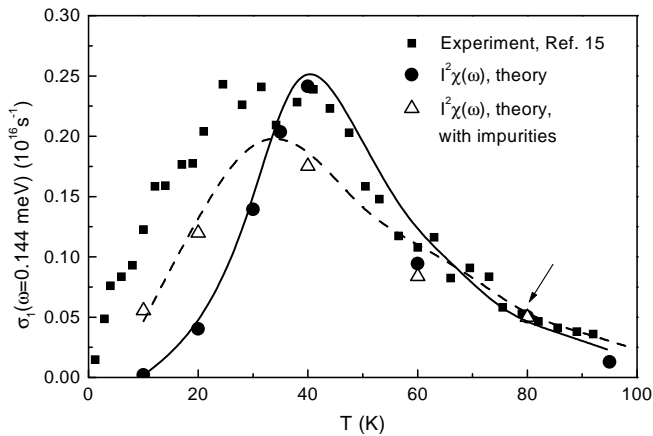}
\caption{Temperature dependence of the conductivity
$\sigma_1(\omega)$ at microwave frequency $\omega = 0.144\,$meV.
The solid circles are results based on our model spectral density
with the solid curve a guide to the eye.
The open triangles include impurities with the dahsed line
to guide the eye. The solid squares
represent experimental data by Bonn {\it et al.} \cite{Bonn1}.}
\label{f4}
\end{figure}
is observed in $\sigma_1(\omega)$ for $\omega = 0.144\,$meV around
$30\,$K. This peak has been attributed to the collapse of the
inelastic scattering as $T$ is lowered in the
superconducting state. If the important scattering has its origin
in correlations effects, as it does in a spin fluctuation mechanism,
it is expected to be strongly affected by the onset of superconductivity,
thus $I^2\chi(\omega)$ should be gapped.
This mechanism is already included in our work with the spin resonance
determining the low $\omega$ part of the spectral density. Our theoretical
results for $\sigma_1(\omega = 0.144\,{\rm meV})$ as a function of
temperature are shown as solid circles in Fig.~\ref{f4}
and are found to be close to previous theoretical results \cite{Schach3}.
The arrow shows
the point at which the theoretical calculations have been made to agree
exactly with the measurements of Bonn {\it et al.} \cite{Bonn1}
(solid squares).
As for the $T$ dependence of $\sigma_1(\omega = 0.144\,%
{\rm meV})$, the agreement with experiment is good at the
higher temperatures, but the theoretical peak is too narrow.
This discrepancy can be removed by including
a small amount of impurity scattering which yields the open triangles.
We note that the temperature dependence of the microwave conductivity
reflects most importantly the reduction to near zero of
$I^2\chi(\omega)$ at small $\omega$ which accompanies the
formation of the resonance peak rather than the peak directly.
Previous work \cite{Schach3} which included a low frequency
cutoff but no resonance peak was equally able to describe the
data and fell close to the solid and dashed lines of Fig.~\ref{f3}.

The plasma frequency $\Omega_p$ which does not enter our theoretical
work can be found by scaling theoretical infrared conductivity data
to experiment \cite{Schach5}. A value of $2.36\,$eV is found (see
Table~\ref{t1}) which compares well with experiment.

A further comparison of our model with the infrared data is provided
by the analysis of the fraction of the total normal state spectral
weight which condenses into the superfluid: $n_s/n$. Indeed, strong
electron-boson coupling reduces the spectral weight of the
quasiparticle component of the electronic spectral function
$A({\bf k},\omega)$ compared to its non-interacting value by
a factor of $Z$ leading at the same time to the appearance of an
incoherent component. It is the latter component which is
responsible for the Holstein band in the optical conductivity
whereas the coherent  quasiparticle part gives rise to the
Drude term at $T>T_c$ and to superfluid density at $T=0$ in
the spectra of $\sigma_1(\omega)$ \cite{footn1}. The values
of $n_s/n$ and hence $(Z-1)$ yield an estimate of the strength of
renormalization effects in the interacting system.
Tanner {\it et al.} \cite{Tanner} obtained $n_s/n\simeq 0.25$
in crystals of Y123 and Bi2212. This compares well with the value
$\simeq 0.33$
which corresponds to $Z\simeq 3$ (at low temperatures) generated
in our analysis. The resonance peak alone accounts for 75\%
of the renormalization effect.

We have also calculated the condensation energy \cite{Carb2} as a function
of temperature. Its value at $T=0$
follows from the normal state electronic density of states which we take
from band structure theory equal to $2.0\,$states/eV/Cu-atom
(double spin) around the middle of the calculated range of values
\cite{Junod}.
This gives a condensation energy $\Delta F(0) = 0.287\,{\rm meV/Cu-atom}$
which agrees well with the value quoted by Norman {\it et al.} \cite{Norm}
from the work by Loram {\it et al.} \cite{Loram}. (See Tab.~\ref{t1}.)
This is equivalent to a thermodynamic critical field
$\mu_0 H_c(0) = 1.41\,$T with $H_c(T)$ defined through
$\Delta F(T) = H^2_c(T)/8\pi$. The normalized value $H_c(T)/H_c(0)$ is shown
as the dotted line in the bottom frame of Fig.~\ref{f3} and is
seen to follow reasonably, but not exactly, the $T$
dependence of the normalized penetration depth. Another quantity
that comes out directly from our calculations is the temperature
dependence of the gap. It follows closely the temperature dependence of the
resonance intensity $A(T)/A(T=10\,{\rm K})$ as
shown (dashed curve) in the top frame of Fig.~\ref{f3}.
One further quantity is the ratio of the gap amplitude
to the critical temperature which in BCS theory is
$2\Delta_0/k_BT_c = 4.2$. In Eliashberg theory the gap depends
on frequency. In this case an unambiguous definition
of what is meant by $\Delta_0$ is to use the position in energy
of the peak in the quasiparticle density of states which
is how the gap $\Delta_0$
is usually defined experimentally for a $d$-wave superconductor. We get
a theoretical value of $ 2\Delta_0/k_BT_c \simeq 5.1$ in good agreement
with experiment, as shown in Tab.~\ref{t1}.
\begin{table}
\caption{Some superconducting properties of the twinned Y123 sample:
$\Delta F(0)$ is the condensation energy at $T=0$ in meV/Cu-atom,
$n_s/n$ is the superfluid to total carrier density ratio, $\Omega_p$ is
the plasma frequency in eV.
\label{t1}}
\begin{center}
\begin{tabular}{cccc}
 & Theory & Experiment & Ref. \\
\hline
$\Delta F(0)$ & 0.287 & 0.25 & \cite{Norm,Loram} \\
$n_s/n$ & 0.33 & 0.25 & \cite{Tanner} \\
$\Omega_p$ & 2.36 & 2.648 & \cite{Tanner2} \\
$2\Delta_0/k_BT_c$ & 5.1 & 5.0 & \cite{Dynes} \\
\end{tabular}
\end{center}
\end{table}

The analysis presented above argues for the prominence of the
spin resonance in the charge dynamics and thermodynamics of
Y123 (and Bi2212) \cite{footn2}. Results from other experimental
techniques which also probe the charge related properties of the
high-$T_c$ superconductors have also been described in terms of
coupling to a collective mode.
ARPES results in Bi2212 \cite{Norm1,Shen} as well
as certain features of tunneling spectra \cite{Chub} are examples.

The analysis of optical data gives the charge carrier-spin
excitation spectral density $I^2\chi(\omega)$ which determines
the superconducting properties of the system within
a generalized $d$-wave Eliashberg formalism. $I^2\chi(\omega)$
depends significantly on $T$ because of feedback
effects expected in theories of electronic mechanisms \cite{Dahm}.
We obtained agreement with experiment for the $T$
dependence of the London penetration depth, of the peak in the
microwave conductivity, and of the spectral weight under the
$41\,$meV spin resonance. The size of the zero temperature condensation
energy is also understood as is the observed value of the
fractional oscillator strength in the condensate and the ratio
of gap to critical temperature.

Research supported in part by NSERC (Natural Sciences and
Engineering Research Council of Canada) and by CIAR (Canadian
Institute for Advanced Research). Work at UCSD is supported through the
NSF grant DMR-9875980. DNB is a Cottrell fellow of the Research
Corporation.


\begin{thebibliography}{99}
\bibitem{footn2}
Y123 and Bi2212 appear to be similar in response
of the CuO$_2$ planes and spin dynamics probed with spin
polarized neutron scattering.
\end{thebibliography}
\end{document}